\documentclass[12pt]{iopart}

\usepackage[dvips]{graphicx}  
\usepackage{color}           
\usepackage{hyperref}       
\usepackage{subfigure}

\begin{document}

\title{Dynamical generation of two-dimensional matter-wave discrete solitons}

\author{Artem M. Dudarev\dag \ddag \footnote[3]{To whom correspondence should be addressed (dudarev@physics.utexas.edu)}, Roberto B. Diener\dag \P and Qian Niu\dag
}

\address{\dag\ Department of Physics, The University of Texas,
Austin, Texas 78712, USA}

\address{\ddag\ Center for Nonlinear
Dynamics, The University of Texas, Austin, Texas 78712, USA}

\address{\P\ Department of Physics, The Ohio State University, Columbus, 
Ohio 43210, USA}

\begin{abstract}
	We suggest a method to experimentally obtain two-dimensional matter-wave discrete solitons with a {\it self-repulsive} BEC in optical lattices. At the edge of the Brillouin zone, a wave packet effective mass is negative which could be treated as inversion of the nonlinearity sign. Above critical nonlinearity this makes the wave packets collapse partially into localized modes with a chemical potential located in the gap between the first and the second bands. This critical nonlinearity is also associated with the smallest nonlinearity for which the discrete solitons are possible in the gap. Extensive numerical simulations for square and asymmetric honeycomb lattices in continuous model illustrate every stage of the process.
\end{abstract}

\maketitle

\section{Introduction}
Periodic lattices with substantial nonlinearities appear in various systems such as biological molecules~\cite{davydov73}, nonlinear optical wave guides~\cite{christodoulides88}, solid-state materials~\cite{su79,sievers88}, and Bose-Einstein condensates (BEC)~\cite{trombettoni01}. In these systems, interplay between linear coupling effects among adjacent sites and nonlinearity can result in a self-localized state - lattice or `discrete' soliton (DS)~\cite{davydov73,christodoulides88,su79,sievers88,trombettoni01}. Until recently direct observation of DS has been done only in one-dimensional (1D) optical wave guides~\cite{christodoulides88,eisenberg98,morandotti99}. Yet in systems with dimensionality more than one a number of fundamental phenomena, such as vortex lattice solitons, bright lattice solitons that carry angular momentum, are expected~\cite{malomed01}. Recently a novel experimental technique to produce photonic crystal with optical induction allowed the authors of Ref.~\cite{fleischer03} to directly observe two-dimensional (2D) DS. 

Dynamics of the optical pulses in nonlinear photonic crystals and BEC in optical lattices are governed by a nonlinear Shr\"odinger equation (NLSE) with periodic potential, hence many predictions done with respect to photonics are expected with BEC. In the case of BEC, the nonlinear coefficient can be either positive or negative for repulsive or attractive atomic interaction respectively, with most of the experiment being done with repulsive atoms. One-dimensional solitons in the absence of periodic potential were observed in attractive BECs~\cite{strecker02,khaykovich02}. One-dimensional matter-wave DS in optical lattices were studied extensively theoretically both for attractive and repulsive interactions~\cite{carusotto02,louis03,ahufinger03}.  Decoherence of the repulsive BEC during Bloch oscillation in 1D optical lattices observed in experiments~\cite{anderson98,burger01,morsch01} was related theoretically to generation of DS~\cite{trombettoni01,scott03} (similar decoherence phenomena in two- and three-dimensional (3D) optical lattices have been reported~\cite{greiner01,greiner02}). After the submission of this article, the observation of the matter-wave DS was reported in 1D~\cite{eiermann04}.

In contrast to free space, stable localized modes are possible in periodic potentials in any dimension both for attractive and repulsive interaction.
In the case of a {\it self-repulsive} BEC generation of multidimensional matter-wave DS due to a modulational instability has been predicted theoretically~\cite{baizakov02}, and the existence and stability of 2D DS have been studied~\cite{ostrovskaya03}. Using a variational approximation and direct numerical simulation the authors of Ref.~\cite{baizakov03} demonstrated that in the case of {\it attractive} interaction above the threshold number of atoms, the initial BEC wave packet placed in optical lattice collapses into multidimensional DS.

The effect of the lattice for quantum wave packets much larger than the unit cell of the periodic potential, may be replaced by the effective mass. In which case, even for {\it repulsive} interaction the wave function envelope dynamics, for points in momentum space where effective mass is negative, are governed by the NLSE with negative (self-focusing) nonlinearity~\cite{steel98,pu03}. The center of the quantum wave packet in momentum space can be easily shifted in a controlled manner by accelerating the lattice (for instance by chirping the relative detuning of the beams creating the lattice) as was demonstrated in early experiments with cold atoms~\cite{raizen97} and later with BEC~\cite{denschlag02}. Recently, effects of negative effective mass have been studied with $^{87}$Rb condensates in one dimensional optical lattices~\cite{eiermann03}.

In 2D above critical value of the self-focusing nonlinearity wave packet collapses. The nonlinearity in case of BEC determined by number of atoms, scattering length, effective mass in the lattice and frequency of transverse confinement. When wave packet size becomes comparable to the size of a unit cell effective mass approximation breaks down. If the nonlinearity is close to that one for which DS is supported by the band gap, part of the wave function is transferred to DS and part decays into linear waves. This is a general phenomenon for NLSE when a state is prepared sufficiently close to the localized state~\cite{soffer03}.

In this article, we first obtain a criteria for critical value of interaction using variational approximation for the wave packet envelope dynamics~\cite{desaix91,malomed02} and effective mass approximation~\cite{steel98,pu03}. We also show that this critical nonlinearity is associated with the smallest nonlinearity for which the lattice supports DS~\cite{flach97,kalosakas02,baizakov04}, and its value is equal to the only nonlinearity with which a stationary solution is possible for the free space 2D NLSE with the corresponding effective mass. In the following section, we illustrate DS generation with a self-repulsive BEC using two numerical examples: square optical lattice for the parameters considered in~\cite{ostrovskaya03} and asymmetric honeycomb lattice. The latter was originally considered as a system to study effects of Berry curvature in periodic potentials~\cite{diener03}. For both systems, we simulate all the stages of the possible experiments: adiabatic introduction of the lattice, half of the Bloch oscillation and wait period for the wave function to collapse to DS. Finally, we summarize the results of this study.

\section{Variational Approximation}
\label{sec:va}
The variational approximation for the NLSE was originally developed in nonlinear optics (for the review see~\cite{malomed02}). It was successfully applied to describe BEC dynamics~\cite{carr02,abdullaev03,perez-garcia00,tsurumi01,salasnich02}, including its evolution in the optical lattices~\cite{trombettoni01,baizakov03,abdullaev01}. In this section we first apply variational approximation ideas to the evolution of the BEC Gaussian wave packet in free space in $D$ dimensions. 

The Gross-Pitaevskii (GP) equation is a NLSE describing the dynamics of BEC:
\begin{equation}
	\label{eq:gp}
	i\hbar \frac{\partial }{{\partial t}}\psi ({\bf r},t) = \hat H \psi ({\bf r},t) = \left( { - \frac{{\hbar ^2 }}{{2m}}\nabla ^2  + V({\bf r}) + Ng_D \left| {\psi ({\bf r},t)} \right|^2 } \right)\psi ({\bf r},t),
\end{equation}
with a nonlinear term being due to the mean-field treatment of the interaction between the atoms. It can be both positive and negative depending on the scattering length of the atomic collisions. Most of the current experiments deal with self-repulsive BECs (positive scattering length). The wave function in the equation is normalized to unity.

To apply variational approximation, we restrict the dynamics of the quantum wave packet in free space ($V({\bf r}) = 0$) in $D$ dimensions to the form
\begin{equation}
	\Psi _D (r) = \left( {\frac{\alpha }{\pi }} \right)^{D/4} e^{ - (\alpha  + i\beta )r^2 /2},
	\label{eq:ansatz}
\end{equation}
where $\alpha$ and $\beta$ are variational parameters, $\alpha$ being inversely proportional to the width of the wave packet squared.
The semi-classical Lagrangian corresponding to GP equation may be written as
\begin{equation}
	L(\alpha ,\dot \alpha ,\beta ,\dot \beta ) = \left\langle {\Psi _D } \right|i\hbar \frac{\partial }{{\partial t}} - \hat H\left| {\Psi _D } \right\rangle.
\end{equation}
The terms necessary for its calculation are
\begin{equation}
	\left\langle {\Psi _D } \right|i\hbar \frac{\partial }{{\partial t}}\left| {\Psi _D } \right\rangle  = \frac{{\hbar D}}{4}\frac{{\dot \beta }}{\alpha },
\end{equation}
\begin{equation}
	\left\langle {\Psi _D } \right| - \frac{{\hbar ^2 }}{{2M}}\nabla ^2 \left| {\Psi _D } \right\rangle  = \frac{{\hbar ^2 D}}{{4m}}\left( {\alpha  + \frac{\beta ^2 }{\alpha }} \right),
	\label{eq:kin}
\end{equation}
\begin{equation}
	V_{{\rm int} }  = \frac{{N g_D }}{2}\int {dr\left| {\Psi _D (r)} \right|^4  = \frac{{N g_D }}{2}\left( {\frac{\alpha }{{2\pi }}} \right)^{D/2} }.
	\label{eq:int}
\end{equation}
The first term describes temporal evolution, the second is associated with kinetic energy, and last term describes mean field interaction between particles. After the substitution $\gamma  = 1/\alpha$ the Lagrangian equation becomes
\begin{equation}
	\label{eq:disp}
	\gamma \ddot \gamma  - \frac{1}{2}\dot \gamma ^2  = \frac{{2\hbar ^2 }}{{m^2 }} + \frac{{2N g_D }}{m}\gamma \left( {\frac{1}{{2\pi \gamma }}} \right)^{D/2}.
\end{equation}
In the free space case, without mean field interaction $(g_D  = 0)$, this equation gives an exact result for dependence of the wave packet dispersion on time
\begin{equation}
	\label{quadr_disp}
	\gamma^2  = 2(\sigma _0^2  + At^2 ),
\end{equation}
where $\sigma _0^2 = \gamma ^2(t = 0)/2$ is initial spatial second moment of the wave packet, and 
\begin{equation}
	\label{eq:a}
	A = \frac{\hbar ^2}{2 m^2 \gamma ^2(t=0)} = \frac{\hbar ^2}{4 m^2 \sigma _0^2}.
\end{equation}
Equation~(\ref{eq:disp}) is particularly simple in 2D. In this case, the right hand side becomes independent of parameters of the wave packet, and the dynamics is effectively described in the same way as for non-interacting atoms.
The wave packet becomes dispersionless when the right hand side of~(\ref{eq:disp}) vanishes. This gives a criteria for critical interaction strength
\begin{equation}
	\label{eq:g-cr}
	N g_c =  \frac{{2\pi \hbar ^2 }}{ m }.
\end{equation}
For negative nonlinearity, when $\left| g \right| > g_c$, the wave packet collapses.

Quantum motion of a wave packet in periodic potential can be effectively described as motion in free space using the concept of effective mass, which in principle may be negative. We comment on this in the next section.

\section{Effective Mass}
When the external potential $V({\bf r})$ is periodic, and the size of the quantum wave packet is much larger than its unit cell, the effect of the potential for different wave vectors ${\bf k}_0$ may be described in terms of the effective mass. This mass may be inferred from the band dispersion~\cite{steel98,pu03}
\begin{equation}
	\label{eq:m-eff}
	m_{{\rm eff},\mu \nu }  = \hbar ^2 \left( {\frac{{\partial ^2 E}}{{\partial k_\mu  \partial k_\nu  }}} \right)^{ - 1},
\end{equation}
When the wave packet is delocalized in real space over many lattice sites, in momentum space it is localized around a given wave vector ${\bf k}_0$. In this case it may be represented as a sum of envelopes $f _n({\bf r})$ over different Bloch bands
\begin{equation}
	\psi ({\bf r},t) = \sum\limits_n {f_n ({\bf r},t)\phi _{n{\bf k}_0 } ({\bf r})e^{ - iE_{n{\bf k}_0 } t/\hbar } },
\end{equation}
where the envelope $f_n({\bf r},t)$ is a slowly varying function within the unit cell, and each Bloch function $\phi _{n{\bf k}_0}$ is the solution of the linear eigenproblem
\begin{equation}
	\label{eq:lin}
	\left( { - \frac{{\hbar ^2 }}{{2m}}\nabla ^2  + V({\bf r})} \right)\phi _{n{\bf k}} ({\bf r}) = E_{n{\bf k}} \phi ({\bf r}),
\end{equation}
normalized to the area of the unit cell
\begin{equation}
	\int\limits_{{\rm cell}} {d{\bf r}\left| {\phi _{n{\bf k}_0 } ({\bf r})} \right|^2 } = \Omega.
\end{equation}
In the experiments it is possible to prepare wave packets that populate only the lowest band~\cite{raizen97}.
The NLSE for the envelope incorporates the effects of the external potential into effective mass~\cite{steel98,pu03}
\begin{equation}
	\label{eq:gp-eff}
	i\hbar \left( {\frac{{\partial f_n }}{{\partial t}} + {\bf v}_g  \cdot \nabla f_n } \right) = \left( { - \frac{{\hbar ^2 }}{{2m_{{\rm eff},\mu \nu } }}\frac{{\partial ^2 }}{{\partial x_\mu  \partial x_\nu  }} + Ng'_2 \left| {f_n } \right|^2 } \right)  f _n,
\end{equation}
where ${\bf v} _g$ is drift velocity of the wave packet center 
\begin{equation}
	{\bf v}_g  = \frac{1}{m}\left\langle {\phi _{n{\bf k}_0 } } \right|{\bf \hat p}\left| {\phi _{n{\bf k}_0 } } \right\rangle,
\end{equation}
and effective interaction strength $g' _2$ is given by
\begin{equation}
	\label{eq:g-eff}
	g'_2  = \frac{g_2}{\Omega} \int\limits_{{\rm cell}} {d{\bf r}\left| {\phi _{n{\bf k}_0 } ({\bf r})} \right|^4 }.
\end{equation}
In the particular examples we will discuss below, the wave packet will be driven to the point in the Brillouin zone where ${\bf v} _g = 0$, and effective mass tensor is negative in all  directions, hence the envelope dynamics will be governed by NLSE with negative mass. This can be viewed as a NLSE with {\it positive} mass with {\it inverted} sign of nonlinearity, which is clearly seen from the equation for complex conjugate of the envelope function
\begin{equation}
	i\hbar {\frac{{\partial f_n^* }}{{\partial t}}} = \left( { - \frac{{\hbar ^2 }}{{2\left| {m_{{\rm eff},\mu \nu } } \right|}}\frac{{\partial ^2 }}{{\partial x_\mu  \partial x_\nu  }} - Ng'_2 \left| {f_n^* } \right|^2 } \right) f^*_n.
\end{equation}
This equation can be obtained by taking the complex conjugate of~(\ref{eq:gp-eff}) and using absolute value of the mass.

Hence, in this situation, as long as the condition for the effective mass approximation holds, $m$ in the Eq.~(\ref{eq:disp}) should be replaced by $m_{\rm eff}$. In the case, when an effective mass and nonlinearity have opposite signs and $\left| g'_2\right| > g'_{2,c}$, where critical interaction strength is defined in~(\ref{eq:g-cr}), with mass being replaced by effective mass, the wave packet collapses. Notice that in general from~(\ref{eq:disp}) it follows that dispersion of the wave packets for any $D$ when the effective mass is negative is described by the equation with positive mass and inverted sign of nonlinearity. When the size of the wave packet becomes comparable to the lattice spacing the effective mass approximation no longer holds. As we will discuss in the following section, in 2D there is nonlinearity below which the DS are not supported by the bandgap. If the nonlinearity is sufficiently larger than this delocalizing nonlinearity, part of the wave packet decays into DS and part decays into linear waves.

\section{Delocalization\label{sec:del}}

In contrast to 1D, where DS may correspond to arbitrary nonlinearity, in 2D and higher dimensions DS are possible only for nonlinearity above a critical value~\cite{flach97}. The authors of~\cite{kalosakas02} considered the possibility of observing the delocalization transition with matter-wave DS in optical lattices, when an irreversible change from DS to delocalized states is produced for a slow change in the lattice parameters. This delocalizing nonlinearity may be associated with the critical nonlinearity for a Gaussian wave packet to collapse as discussed in Section~\ref{sec:va}.

The concept of the effective mass is not generally applied to the DS since in the middle of the gap the DS is localized within one lattice site. As the chemical potential of the DS comes close to a band of linear states, its space extension increases, hence one may expect that the effective mass approximation becomes applicable. The results of Section~\ref{sec:va} imply that there is only one value of nonlinearity for which the localized modes are supported in 2D free space when the envelopes of the localized modes are approximated by Gaussians. This also can be shown in general from the scaling arguments for 2D NLSE. Indeed, if the normalized wave function $\psi_1(x,y)$ is the solution of 
\begin{equation}
	- \nabla ^2 \psi _1  + \gamma \psi _1^3  = \tilde \mu \psi _1,   
\end{equation}
then the normalized wave function $\psi_2 = B \psi _1(A x, A y)$ is the solution of
\begin{equation}
	- \nabla ^2 \psi _2  + \gamma \psi _2^3  = A^2 \tilde \mu \psi _1.
\end{equation}
This means that in 2D free space there is only one possible value of $\gamma$ for which localized modes can be found for any value of $\tilde \mu$.

Localized wave packets with very large extension correspond to the critical nonlinearity. As their size is reduced, other corrections due to the lattice also start to play a role. The variational approximation gives a clear picture of what happens to initially localized states of NLSE in 2D. It predicts the critical value of nonlinearity above which evolution of the wave packet width changes character. One may expect that the value of $N g_c$ given in~(\ref{eq:g-cr}) is close to the exact one. We confirmed this expectation by performing direct self-consistent numerical simulations based on the effective potential approach suggested in~\cite{baizakov04}. Similar to the self-consistent Hartree-Fock approximation, one may consider a DS to be a localized state in the effective potential created by itself
\begin{equation}
	\label{eq:eff-pot}
	V_{{\rm eff}} ({\bf r}) =  - Ng'_D \left| {\psi ({\bf r})} \right|^2 .
\end{equation}
In the numerical simulation, we started with an {\it arbitrary} nodeless initial wave function, and with the imaginary time evolution, found the ground state of the potential~(\ref{eq:eff-pot}) for the Hamiltonian given by
\begin{equation}
	\hat H_{{\rm eff}}  =  - \frac{{\hbar ^2 }}{{2\left| {m_{{\rm eff}} } \right|}}\nabla ^2  + V_{{\rm eff}} ({\bf r}),
\end{equation}
and used it in the next step of iteration. We found that independent of the initial guess state above the critical value 
\begin{equation}
	Ng'_{{\rm num}}  = \frac{{5.850 {\rm~} \hbar ^2 }}{{\left| {m_{{\rm eff}} } \right|}}, 
	\label{eq:g-del-num}
\end{equation}
the self-consistent procedure resulted in collapsing states with infinite negative energy, while for smaller nonlinearities the states expanded, with energy going to zero. This value differs from the one for an extended Gaussian wave packet to collapse~(\ref{eq:g-cr}) by $\sim10$ percent.

As an alternative method we also reduced the 2D equation to a 1D ordinary differential equation and solved two point boundary value problem ($\psi'(0) = 0$, $\psi(\infty) = 0$) with the shooting method~\cite{press92}. We found that for arbitrary energy, nodeless solitons in free space with arbitrary size are supported only for one value of nonlinearity given by the same value as in~(\ref{eq:g-del-num}).

The direct dynamical simulations discussed in the next section confirm the existance of the critical nonlinearity above which a wave packet with a finite size collapses. Also, using the imaginary time evolution for different values of chemical potential in the gap, we find corresponding nonlinearity for DS in the lattice to exist. The minimum nonlinearity found as a result of this calculation is also in a good agreement with~(\ref{eq:g-cr}) and~(\ref{eq:g-del-num}).

\section{Numerical Simulations}
\subsection{General Remarks}

By choosing appropriate units of length, $L_u$, and mass, $M_u = M_{\rm atom}$, the GP equation~(\ref{eq:gp}) is reduced to a dimensionless form
\begin{equation}
	\label{eq:gp-2}
	i\frac{{\partial \psi }}{{\partial t}} = \left[ { - \frac{1}{2}\nabla ^2  + V_L ({\bf r}) + N g _2 \left| \psi  \right|^2  + {\bf F} \cdot {\bf r}} \right]\psi .
\end{equation}
We also add an external force ${\bf F}$ that in the case of optical lattices may be created by accelerating the lattice, for example, by sweeping the relative frequency of the beams creating the lattice.
In this case, unit of energy of the problem is 
$E_u = \hbar^2/M_u L_u^2$ and the unit of time is given by $t_u = \hbar / E_u$. 
When the unit of length is chosen to be equal to the inverse wave vector of the light creating the optical lattice
$L_u = 1/k_L$, typical maximum depths of the optical potentials achievable in the dissipationless regime are $\sim 20$. The forces created by accelerating the lattice are limited due to the finite lifetime of the excited states, in case of alkali atoms to about $1000$, in the experiments~\cite{raizen97} forces on the order of $\sim 1-10$ were used.
The dynamics of the BEC is described by a 2D equation in the case when the strong confinement in the transverse direction ``freezes'' the wave function in that direction to the harmonic oscillator ground state. This happens when oscillator length in that direction $l_z = \sqrt{\hbar/M_{\rm atom} \omega_z}$ becomes smaller than the condensate healing length $\xi = \left( 4\pi n a \right)^{-1/2}$, where $n$ is atom density and $a$ is scattering length. When, at the same time, $l_z$ is still larger than 3D scattering length, $\left| a \right|$,
\begin{equation}
	\left| a \right| < l_z  < \xi  \\ 
\end{equation}
the collisions between the atoms preserve their 3D character, yet the dynamics of the BEC in the two other directions are effectively described by 2D GP equation~\cite{petrov00}. Such a regime was recently demonstrated experimentally~\cite{gorlitz01}. The nonlinear coefficient in this case is given by
\begin{equation}
	g_{{\rm 2}}  = \left( {\frac{{8\pi \omega _z \hbar ^3 }}{M_{\rm atom}}} \right)^{1/2} a.
\end{equation}
The experimental system, therefore, provides great flexibility with which the nonlinear coefficient in corresponding NLSE can be controlled. Parameters variable in experiments are number of atoms, $N$, the frequency of transverse confinement $\omega _z$, and the scattering length, $a$, that in principle may be tuned by a magnetic field with Feshbach resonances~\cite{moerdijk95}. For the experimental parameters of~\cite{gorlitz01} we estimate the nonlinear coefficient in present units to be $N g_2 \sim 6000$. For the ansatz chosen above to be a good choice, kinetic energy~(\ref{eq:kin}) should be larger than interaction energy~(\ref{eq:int}). It means that the nonlinearity should be smaller than $\sim 2\pi$. Below we consider only the cases when this holds. For BEC the effective nonlinearity can be always reduced either by changing number of atoms or trapping frequency in transverse direction.

The stationary states of Eq.(\ref{eq:gp-2}) are described by solutions in the form: $\psi ({\bf r},t) = \phi({\bf r}) \exp(-i \mu t)$, where $\mu$ is the chemical potential. Localized states can be found for $\mu$ in the gaps of linear problem.

Below, we consider two examples. The first one is based on parameters for which existence and stability of the solitons were studied in~\cite{ostrovskaya03}. Our consideration extends the treatment to suggest a specific approach for DS generation based on the ideas outlined in previous sections. The second example deals with the asymmetric honeycomb lattice that was studied by us~\cite{diener03} in the context of observing self-rotation and Berry curvature effects for quantum wave packets in asymmetric periodic potentials. There,  robust spontaneous generation of the DS above a critical interaction strength was observed for wave packets left at the corner of the Brillouin zone. Here, we discuss how this effect could be naturally explained in terms of the effective mass concept.

In both cases we have performed simulations in a form that mimics possible experiments. We start with a Gaussian wave packet in free space with a size that is much larger than the unit cell of the potential. The process may be divided into three stages: (1) adiabatic introduction of the lattice potential, (2) acceleration of the lattice for half of the Bloch oscillation, and (3) a wait period for the wave packet to collapse. In the first two stages, adiabaticity is crucial. An adiabatic criteria can be estimated based on Landau-Zener formalism for tunneling between two levels with an avoided crossing. In this case the probability to tunnel is given by~\cite{landau32,zener32}
\begin{equation}
	\label{eq:lz}
	P = \exp \left( - \frac{\pi \delta^2}{2 \alpha} \right),
\end{equation}
where $\delta$ is the gap between the states and $\alpha$ is the rate of change of the gap. This criteria, when a periodic potential with amplitude $V_0$ is introduced in time $t_V$, becomes 
\begin{equation}
	\label{eq:tv}
	\frac{{\delta ^2 t_V }}{{V_0 }} \gg 1,
\end{equation}
where $\delta$ is the gap between the first and the second bands at $k = 0$, since originally the width of the wave packet in the momentum space is much smaller than the size of the Brillouin zone (note that units with $\hbar = 1$, are used).
The maximum ``force'' of the drive may be similarly estimated as
\begin{equation}
	\frac{{\epsilon _g }}{{\frac{{\partial \epsilon }}{{\partial k}}F}} \gg 1.
\end{equation}
Here $\epsilon _g$ is the smallest gap between the first and the second bands, and $\frac{{\partial \epsilon }}{{\partial k}}$ is the largest slope of the dispersion. An alternative expression may be derived based on the WKB approximation~\cite{marder00}
\begin{equation}
	\label{eq:f}
	\frac{{\epsilon _g^{3/2} m_{\rm eff}^{1/2} }}{F} \gg 1  
\end{equation}
where the effective mass, $m _{\rm eff}$, can be estimated from the dispersion as well.
We made sure that the adiabaticity criteria are fulfilled in the simulations discussed below. We also checked it numerically by increasing the driving force and observing the break down of the adiabaticity.

\subsection{Square Lattice}

As a first example, we discuss the model potential considered in~\cite{ostrovskaya03} for existence and stability of 2D DS in continuous potentials, namely
\begin{equation}
	\label{eq:pot-square}
	V_L(x,y) = V_0 (\sin ^2 x + \sin ^2 y).
\end{equation}
Such a potential may be experimentally obtained by overlapping two pairs of counter-propagating beams, far detuned from atomic resonance to reduce effects of spontaneous emission. As in~\cite{ostrovskaya03}, we use amplitude $V_0 = 5$.

\begin{figure*}[floatfix]
	\begin{center}
		\subfigure[]{
		\label{fig:dispersion}
		\includegraphics[width=2.75in]{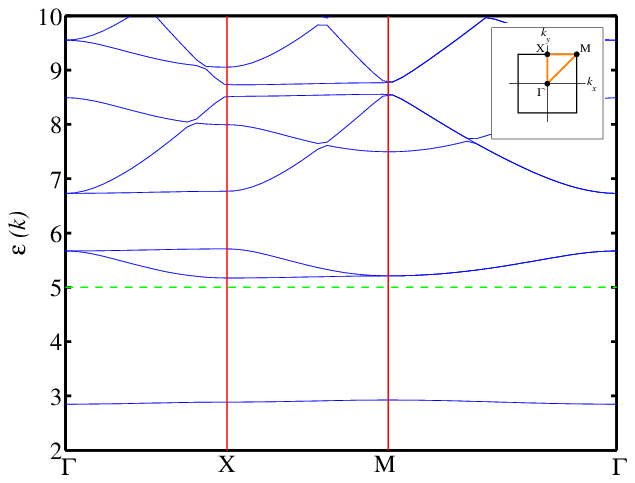}
		}
		\subfigure[]{		
		\label{fig:meff}
		\includegraphics[width=2.75in]{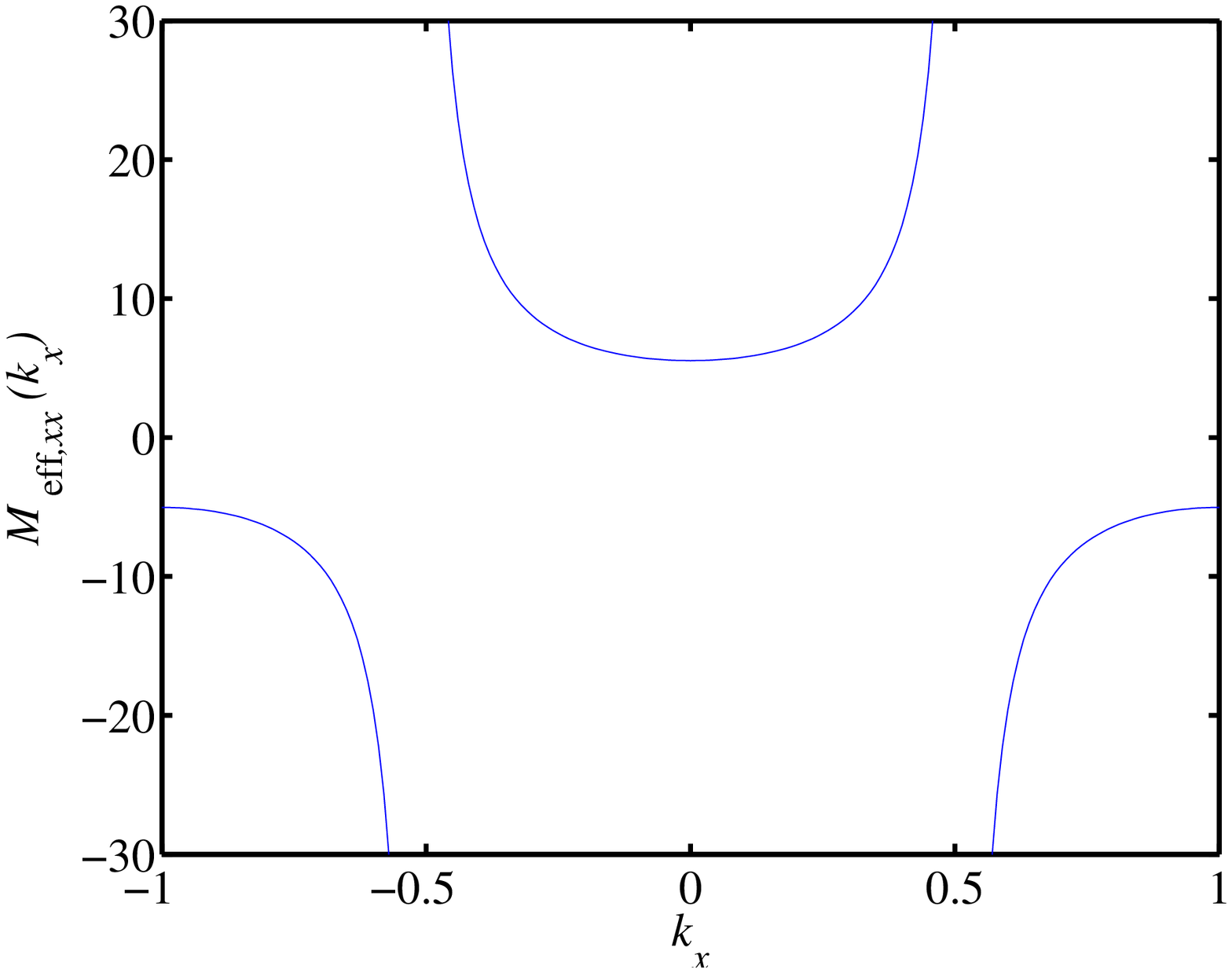}
		}
		\subfigure[]{
		\label{fig:sigma}
		\includegraphics[width=2.75in]{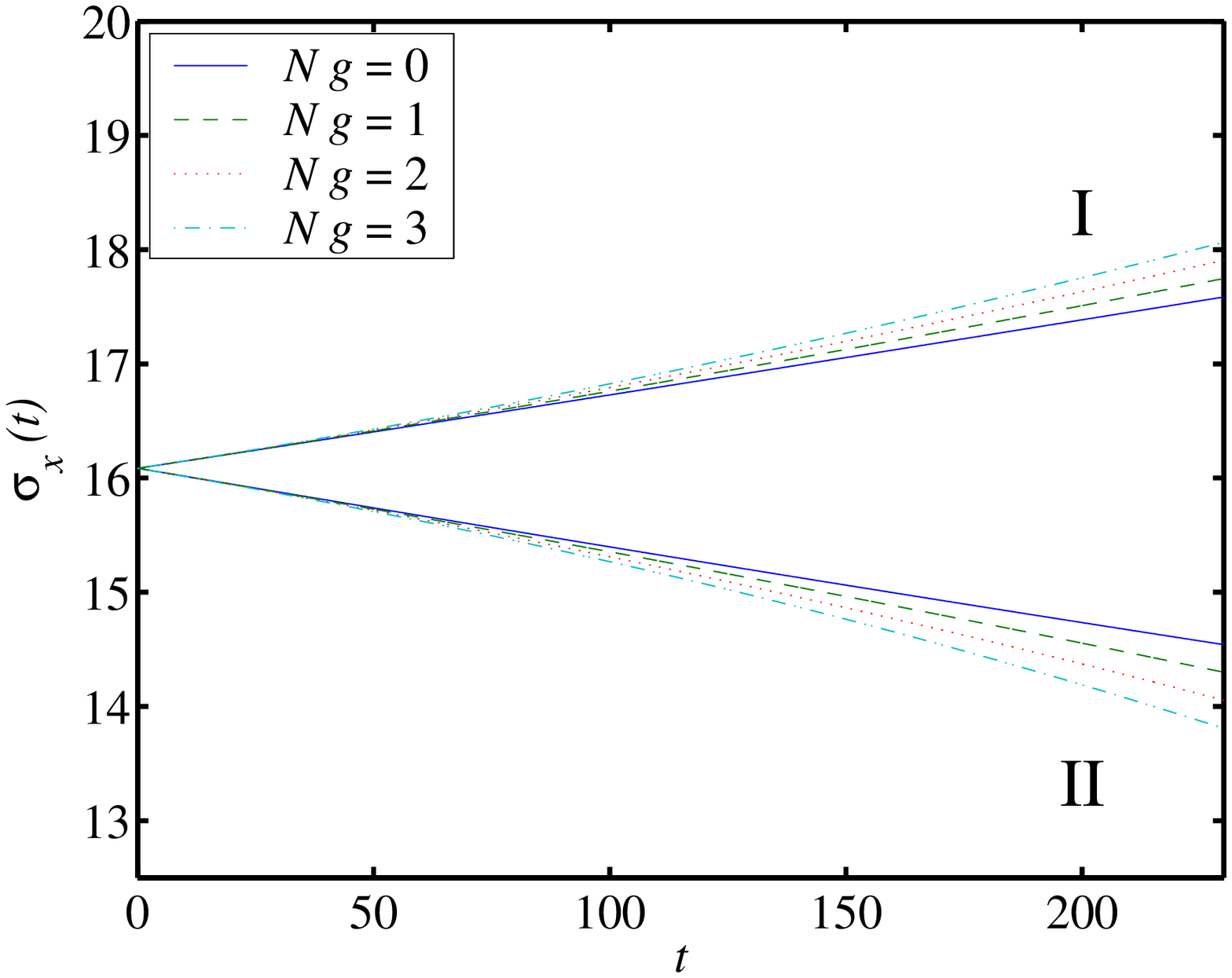}
		}
		\subfigure[]{		
		\label{fig:chi}
		\includegraphics[width=2.75in]{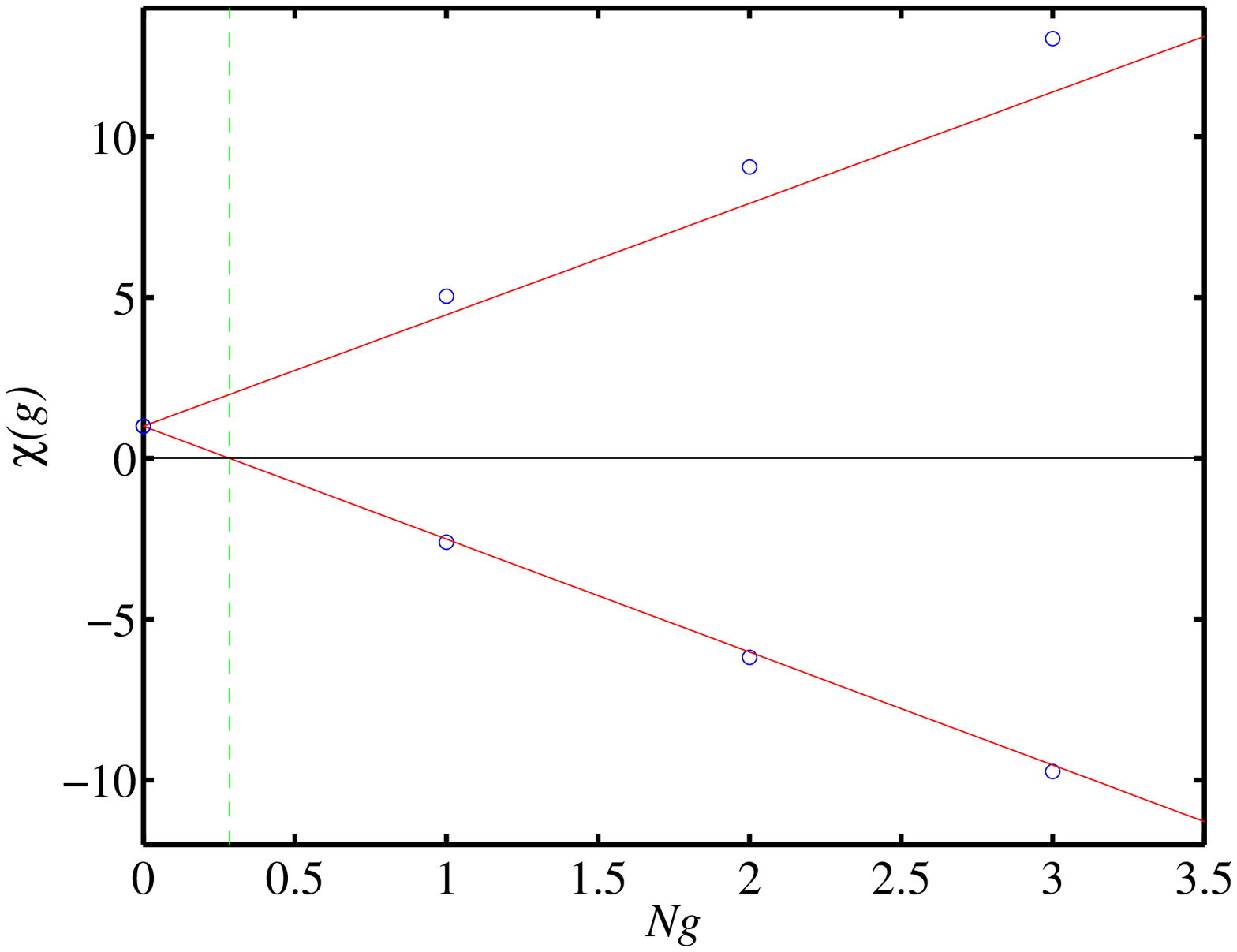}
		}
		\label{fig:sigma_chi1}
		\caption{(a) - Dispersion for square lattice potential~(\ref{eq:pot-square}) along the high symmetry path. The green dashed line shows the boundary between strongly bound and quasi-unbound states~\cite{ostrovskaya03}.
		(b) - $xx$ component of the effective mass tensor along $k_x$-axis. 
		(c) - Dependence of wave packet dispersion on time after the lattice potential is introduced. The two sets of curves are for different points in Brillouin zone: (I) - point $\Gamma$, (II) - point M.
		(d) - Dispersive characteristic $\chi$ from~(\ref{eq:chi2}). The blue dots are obtained from fitting quadratic dispersion~(\ref{quadr_disp}) to continuous simulation data from panel (c), red line is expected behavior of $\chi$.
		}
		\label{fig:dispersion-meff-sigma-chi}
	\end{center}
\end{figure*}

In Fig.~\ref{fig:dispersion}, we show the dispersion of the linear problem~(\ref{eq:lin}) for the potential~(\ref{eq:pot-square}) along the high symmetry directions. The first band has a minimum at zero momentum ($\Gamma$ point), and a maximum at the corner of the Brillouin zone (M point). There exist directions on the plane for which the maximum value of the potential is smaller than $V_0$. These are orthogonal ($x$,$y$) directions, while the absolute maximum of the potential is $2 V_0$. For the chemical potential of the localized states above $V_0$ (shown as a green dashed line in Fig.~\ref{fig:dispersion}), the BEC states are quasi-unbound. For values of $\mu$ close to this boundary an adequate description of DS is not possible within one-band tight-binding model. This is because situations when nodes of the solitonic wave functions are located at the potential minima are possible~\cite{ostrovskaya03}.

The $xx$-component of the effective mass tensor~(\ref{eq:m-eff}) is shown in Fig.~\ref{fig:meff}. Since the potential is separable and symmetric, the $yy$-component has the same dependence on $k_y$, and the tensor is diagonal. At the points of global maximum and minimum of the dispersion, the effective mass in both directions is the same. At the $\Gamma$ point, the effective mass has the smallest positive value  ($m_{{\rm eff},\Gamma} = 5.53$), while at the point M it is negative and has the smallest absolute value ($m_{{\rm eff},{\rm M}}=-5.03$). Hence the smallest nonlinearity is necessary for the wave packet to self-collapse at the M point. The fact that in 1D, for a sinusoidal potential for wave vectors, $k$, larger than half of the largest vector in the Brillouin zone, $k> k_{\rm crit} = \pi / (2 a)$, where $a$ is lattice spacing, allows to give a physical explanation of the origin of the Landau instability studied in~\cite{wu03} in terms of tight-binding approximation. For any interaction strength, wave packets composed of Bloch waves for $k < k_{\rm crit}$ remain wave packets composed of Bloch waves, while for $k > k_{\rm crit}$, when interaction is large enough, they partially collapse to localized modes.

To investigate the validity of the effective mass description, we checked in numerical simulations the predictions that may be based on the formulas discussed in previous sections. From~(\ref{eq:a}), we introduce the quantity characterizing dispersion of the wave packet 
\begin{equation}
	\label{eq:chi}
	\chi  = \frac{{A\sigma ^2 }}{{A\left. {\sigma ^2 } \right|_{g_2  = 0} }} = 1 + \frac{{Ng'_2 m_{{\rm eff}} }}{{2\pi }}, \\ 
\end{equation}
which changes linearly with interaction. As it becomes negative,  the wave packet collapses. Notice that in this expression the effective interaction strength is related to continuous interaction strength with~(\ref{eq:g-eff}), which for the case of separable sinusoidal potential at least partially may be computed analytically. For the separable potential~(\ref{eq:pot-square}) the solution of the stationary eigenproblem~(\ref{eq:lin}) is separable: $ \phi (x,y) = \phi _x (x)\phi _y (y) $, where each wave function is given by the solution of the corresponding equation
\begin{equation}
	\label{eq:eig-1d}
	\left[ {\frac{1}{2}\frac{{\partial ^2 }}{{\partial x^2 }} + \left( {E - \frac{{V_0 }}{2}} \right) + \frac{{V_0 }}{2}\cos 2x} \right]\phi _x (x) = 0.
\end{equation}
Due to Bloch-Floquet theorem, the solution is a product of a periodic Bloch function $u_k(x)$ and a plane wave
\begin{equation}
	\psi _{x,k_x}(x) = e^{ik_x x}u_k(x).
\end{equation}
After the following substitutions 
\begin{equation}
	\begin{array}{c}
		b = 2\left( {E - \frac{{V_0 }}{2}} \right), \\
		z = x, \\ 
		q =  - \frac{{V_0 }}{2}, \\ 
		\psi _x (x) = y(z), \\
	\end{array}
\end{equation}
Eq.~(\ref{eq:eig-1d}) becomes the Mathieu equation~\cite{blanch64}
\begin{equation}
	\label{eq:mathieu}
	\frac{{\partial ^2 y}}{{\partial z^2 }} + (b - 2q\cos 2z)y = 0,
\end{equation}
as a result the solution of~(\ref{eq:eig-1d}) are Mathieu functions with characteristic value $b$ and parameter $q$ 
\begin{equation}
\psi _{x,k_x } (x) = {\rm Ce}(b,q,z) = {\rm Ce}\left( {2\left( {E - \frac{{V_0 }}{2}} \right), - \frac{{V_0 }}{2},x} \right).
\end{equation}
Introducing the numerical factor 
\begin{equation}
	I = \frac{{\left( {\int\limits_{ - \pi /2}^{\pi /2} {Ce_r^4 } (a,q,z)dz} \right)^2 }}{{\left( {\int\limits_{ - \pi /2}^{\pi /2} {Ce_r^2 } (a,q,z)dz} \right)^4 }} 
\end{equation}
we get the following dependence of $\chi$ on interaction strength, 
\begin{equation}
	\label{eq:chi2}
	\chi  = 1 + \frac{{\Omega m_{{\rm eff}} I}}{{2\pi }}Ng_2.
\end{equation}
Therefore the critical interaction strength is given by
\begin{equation}
	Ng_{2,c}  = \frac{{2\pi }}{{\Omega \left| {m_{{\rm eff}} } \right|I}}.
\end{equation}
For the considered example, area of the lattice unit cell is $\Omega = \pi ^2$ and the numerical factor for the top of the band is $I _{\rm M}= 0.398$, which makes the critical interaction strength $N g_{2,c} = 0.285$.

\begin{figure}[floatfix]
	\begin{center}
		\subfigure[]{
		\label{fig:ng-vs-mu}
		\includegraphics[width=2.75in]{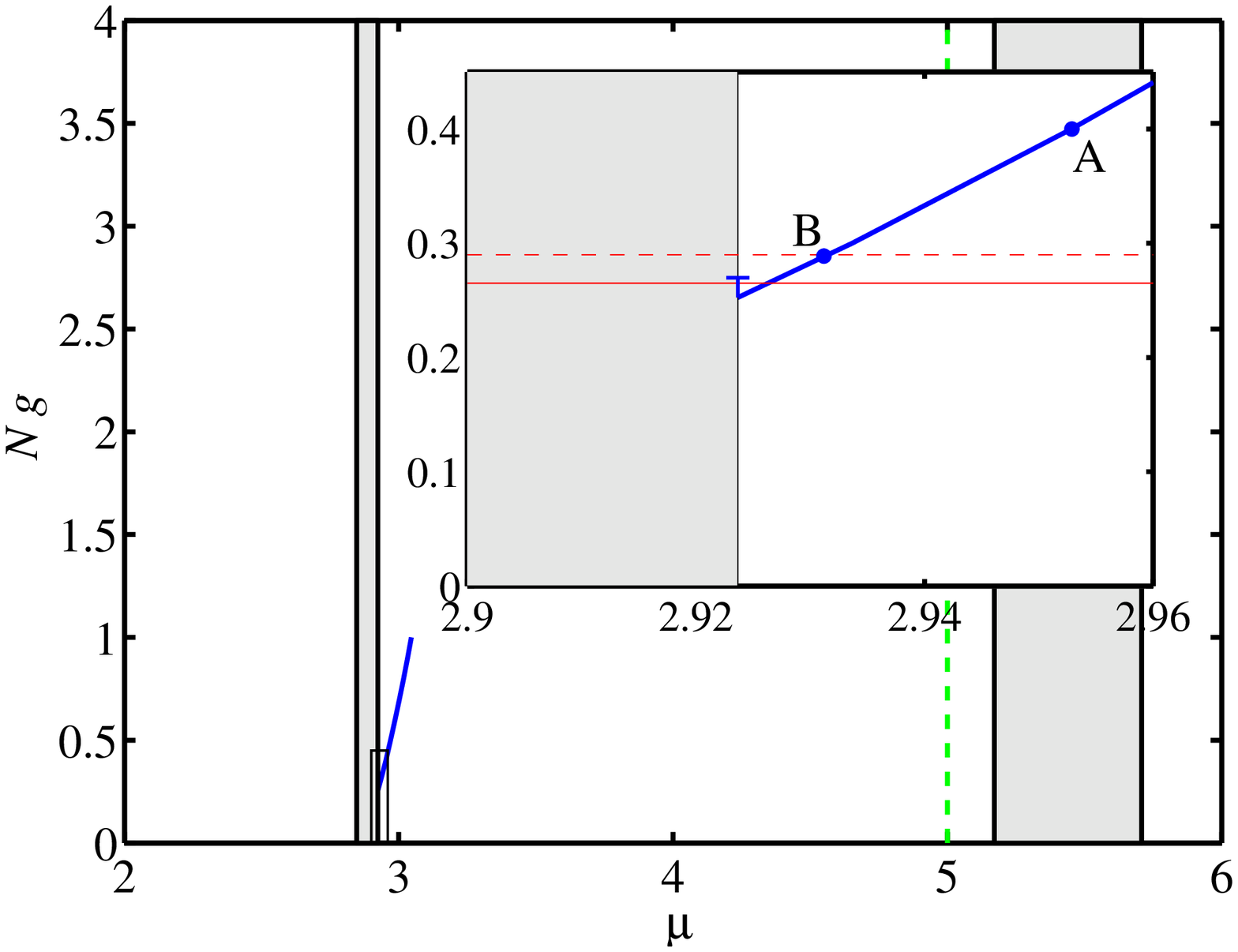}
		}
		\subfigure[]{
		\label{fig:soliton-analit}
		\includegraphics[width=2.75in]{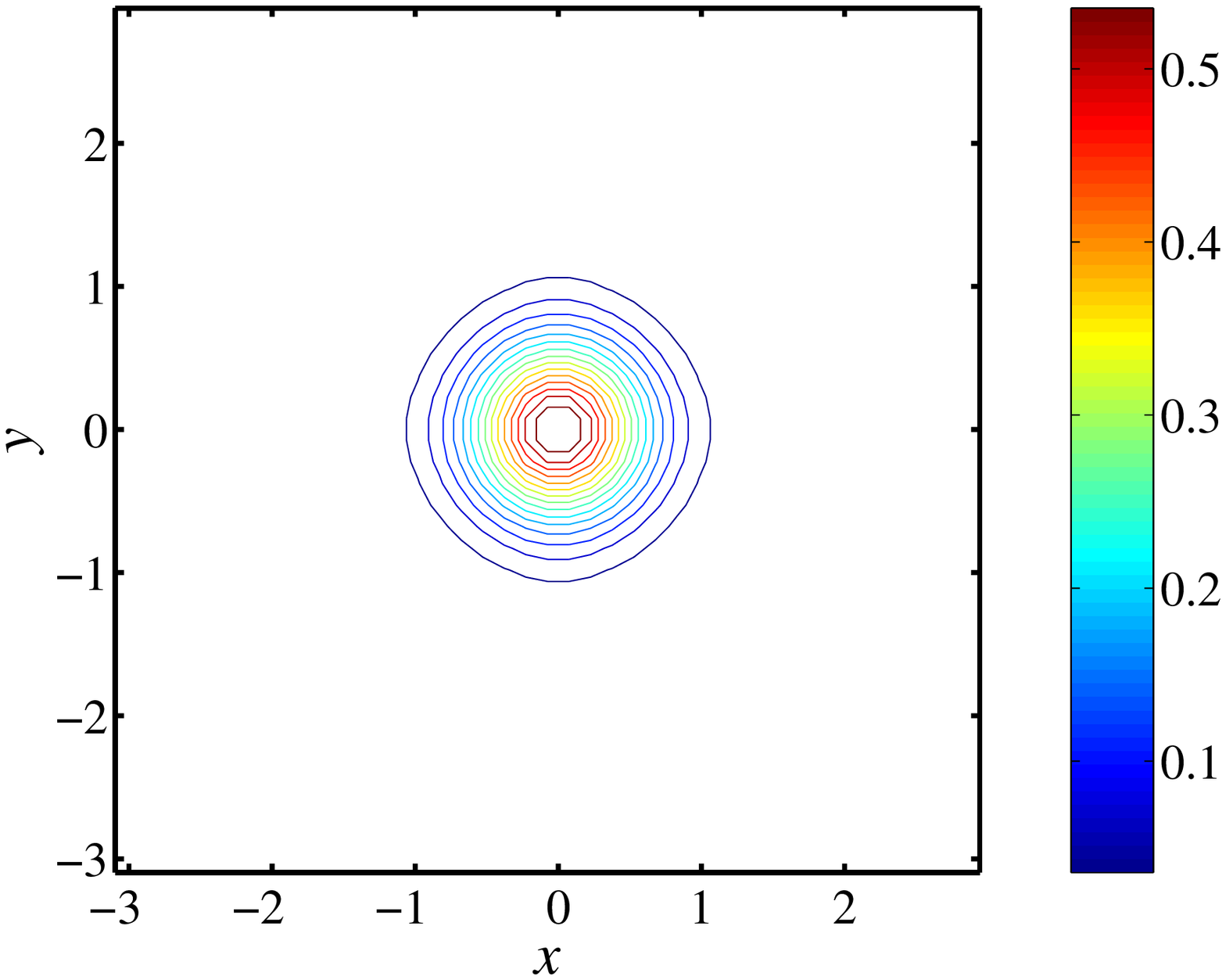}
		}
		\subfigure[]{
		\label{fig:c02}
		\includegraphics[width=2.75in]{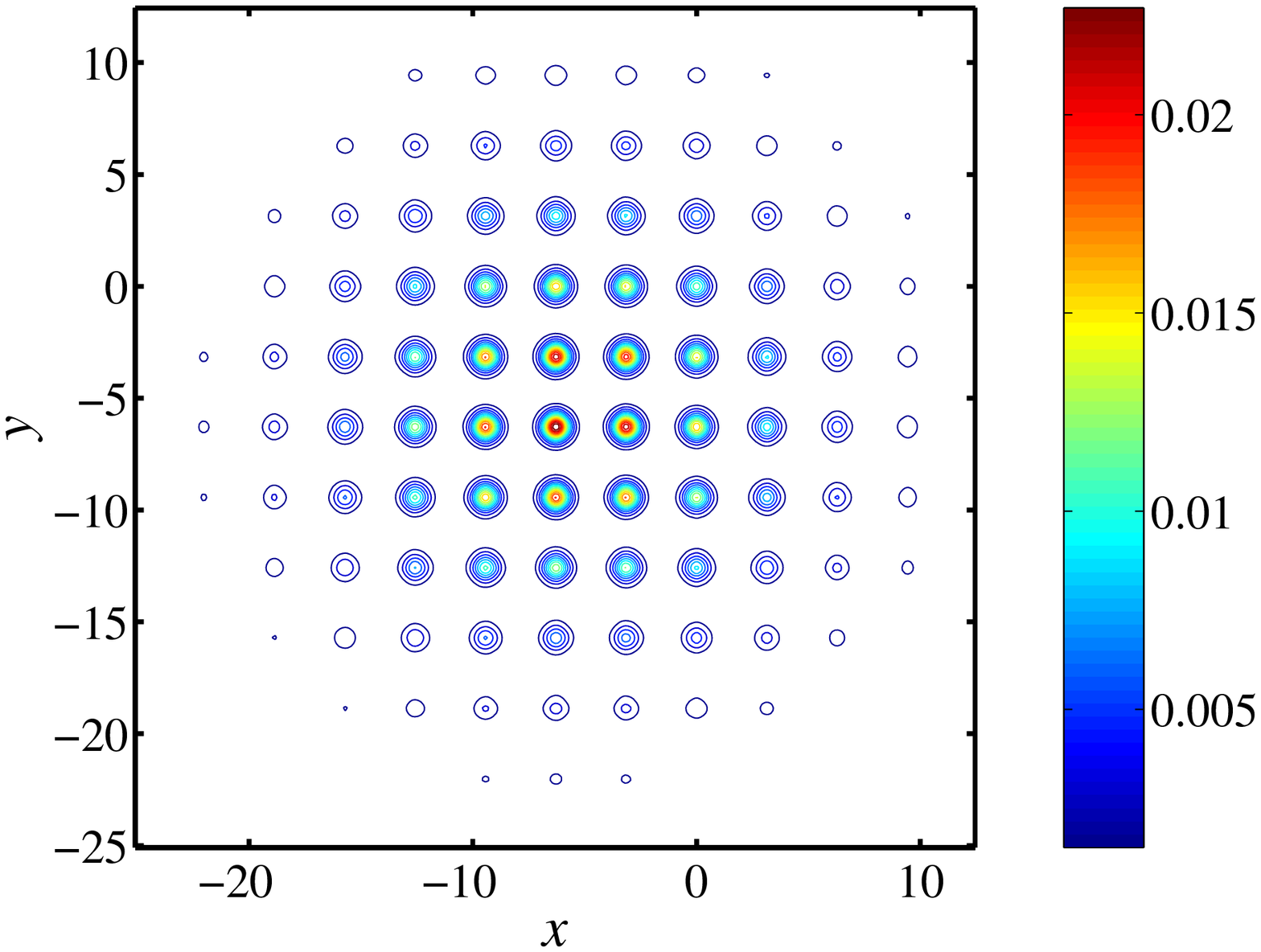}
		}
		\subfigure[]{
		\label{fig:c04}
		\includegraphics[width=2.75in]{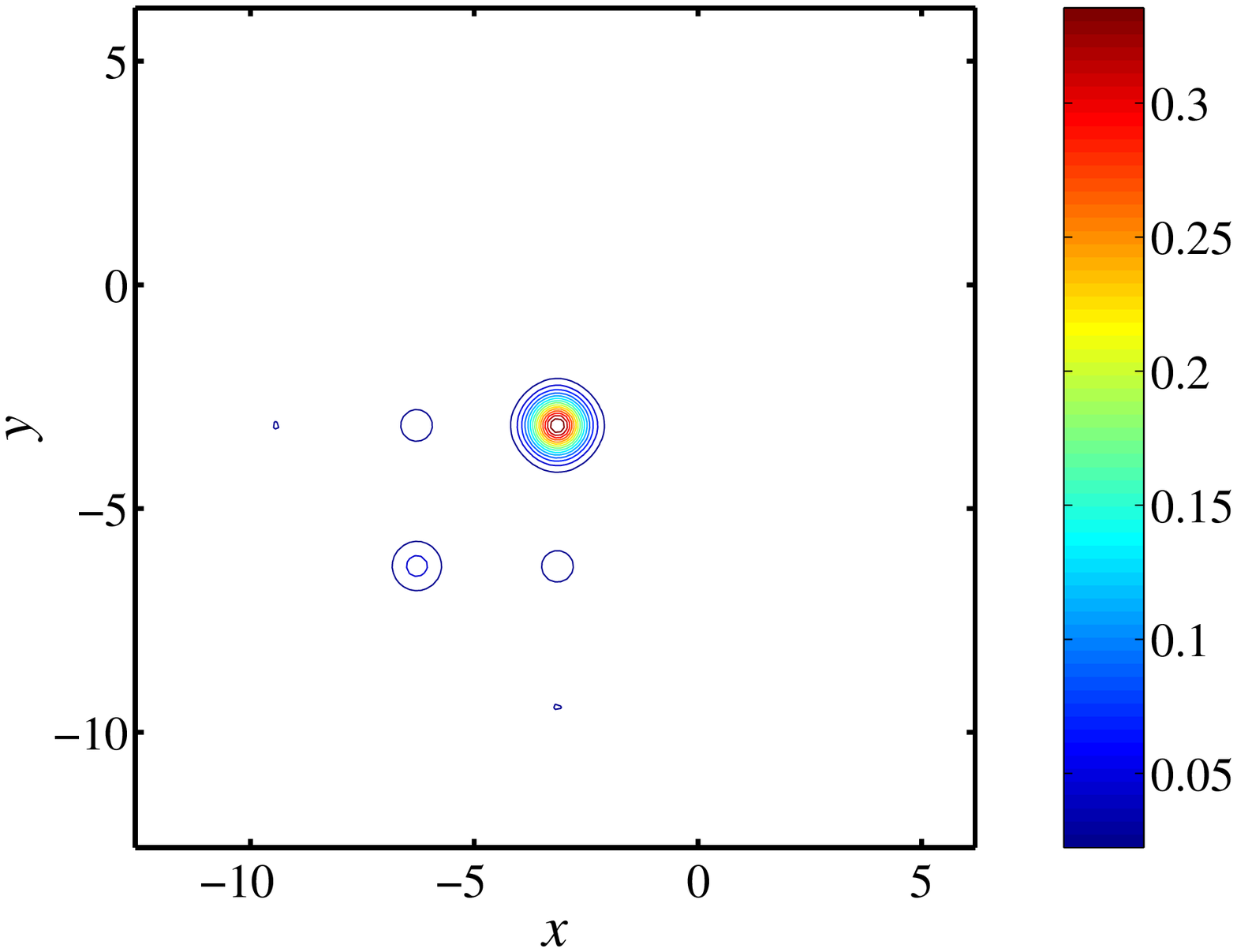}
		}		
		\caption{
		(a) - Dependence of nonlinearity on corresponding chemical potential. The first and second bands are shown with solid rectangles. The inset shows a region denoted by a rectangle in the lower left corner. Horizontal solid line shows numerical value for delocalizing nonlinearity, $g_{\rm num}$, dashed line is the critical nonlinearity for extended Gaussian wave packet to collapse $g_{2,c}$.
		(b) - Spatial distribution of DS corresponding to $\mu = 0.7$ obtained with the descent method.
		(c) - Probability distribution for BEC wave function evolved with $N g_2 = 0.2$ for $\Delta t \sim 1500$ after it was driven to M point.
		(d) - The same for $N g = 0.4$ (point A in (a)), approximately 0.72 of the wave function probability is transferred to the soliton, which corresponds to an effective nonlinearity of $N g_{2,{\rm eff}} \approx 0.288$ - at the top of the first band (point B in (a)).}
		\label{fig:ng-solitons}
	\end{center}
\end{figure}

To test predicted behaviour of the parameter $\xi$, we perform a numerical simulation with a continuous potential. We start with a wave packet of size $\sigma _{x,y} = 15/\sqrt{2}$, while the size of the unit cell is $\pi$ in each direction. The lattice potential is ramped from $V_0 = 0$ to $V_0 = 5$ in $t_V = 450$, then the lattice is accelerated with force $F=0.01$ in diagonal $[11]$ direction. The acceleration is halted after the wave packet undergoes half of the Bloch oscillation, in other words, when its center is located at point M in momentum space. After this, the wave packet expands freely in the presence of the lattice potential. Ramp time and the force satisfy adiabaticity conditions~(\ref{eq:tv})-(\ref{eq:f}), so that the wave packet during evolutions stays in the first band.  In addition to expanding the wave packet at the top of the band, we have studied the expansion as soon as the lattice was introduced without introducing external field, {\bf F}. In both cases the agreement with analytic prediction~(\ref{eq:chi}) is very good (see Fig.~\ref{fig:sigma} and Fig.~\ref{fig:chi}).

As the external force is removed the wave packet contracts due to phases accumulated during the acceleration even in linear case. But this is only when the interaction is above critical that the localized modes are formed. To find a stationary localized solution we followed optimization procedure based on a descent technique with Sobolev preconditioning used in~\cite{ostrovskaya03} and described in~\cite{ripoll01}. In Fig.~\ref{fig:ng-vs-mu} we display dependence of the nonlinearity on chemical potential of the DS. The error bar shows the estimated uncertainty for the curve to intersect the band of extended states obtained by interpolating to infinite size and infinite relaxation time of the descent procedure. The agreement with the argument based on effective mass and free space solitons given in Section~\ref{sec:del} is excellent. As one starts with an extended wave packet and nonlinearity supported by the gap, it shrinks and may lose some part to radiation (extended states), so that the effective interaction experienced by the localized mode is then just a fraction of the actual interaction. As shown in Fig.~\ref{fig:ng-solitons}, the solitons are formed inside the gap. When nonlinearity increases, their chemical potential increases but not significantly, so that they stay relatively close to the top of the first band.

\subsection{Asymmetric Honeycomb Lattice}

\begin{figure*}[floatfix]
	\begin{center}
		\subfigure[]{
		\label{fig:dispersion_h}
		\includegraphics[width=2.75in]{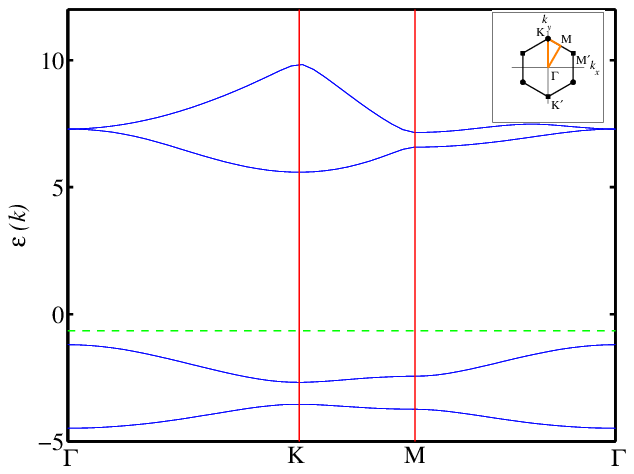}
		}
		\subfigure[]{		
		\label{fig:meff_h}
		\includegraphics[width=2.75in]{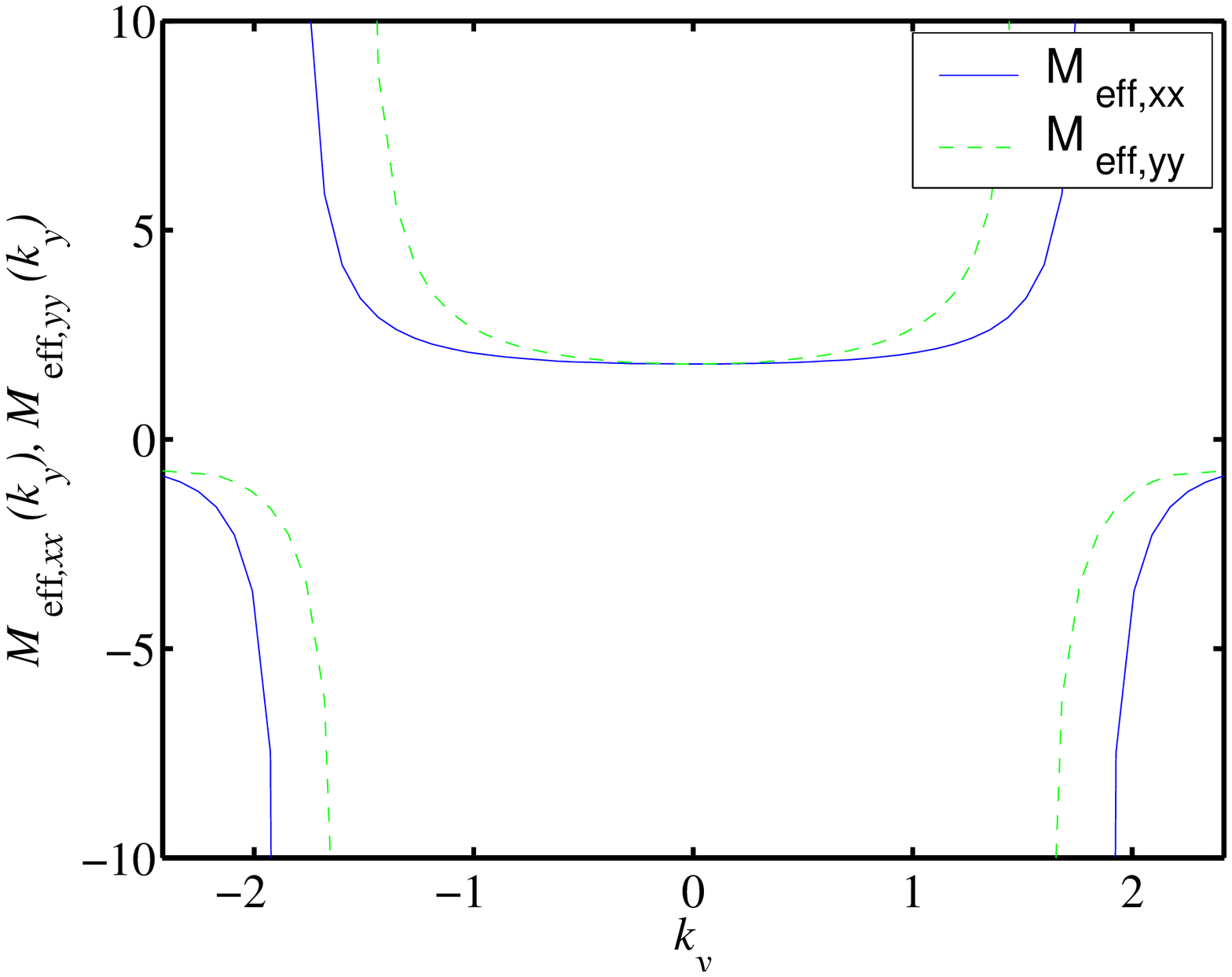}
		}
		\subfigure[]{
		\label{fig:sigma1_h}
		\includegraphics[width=2.75in]{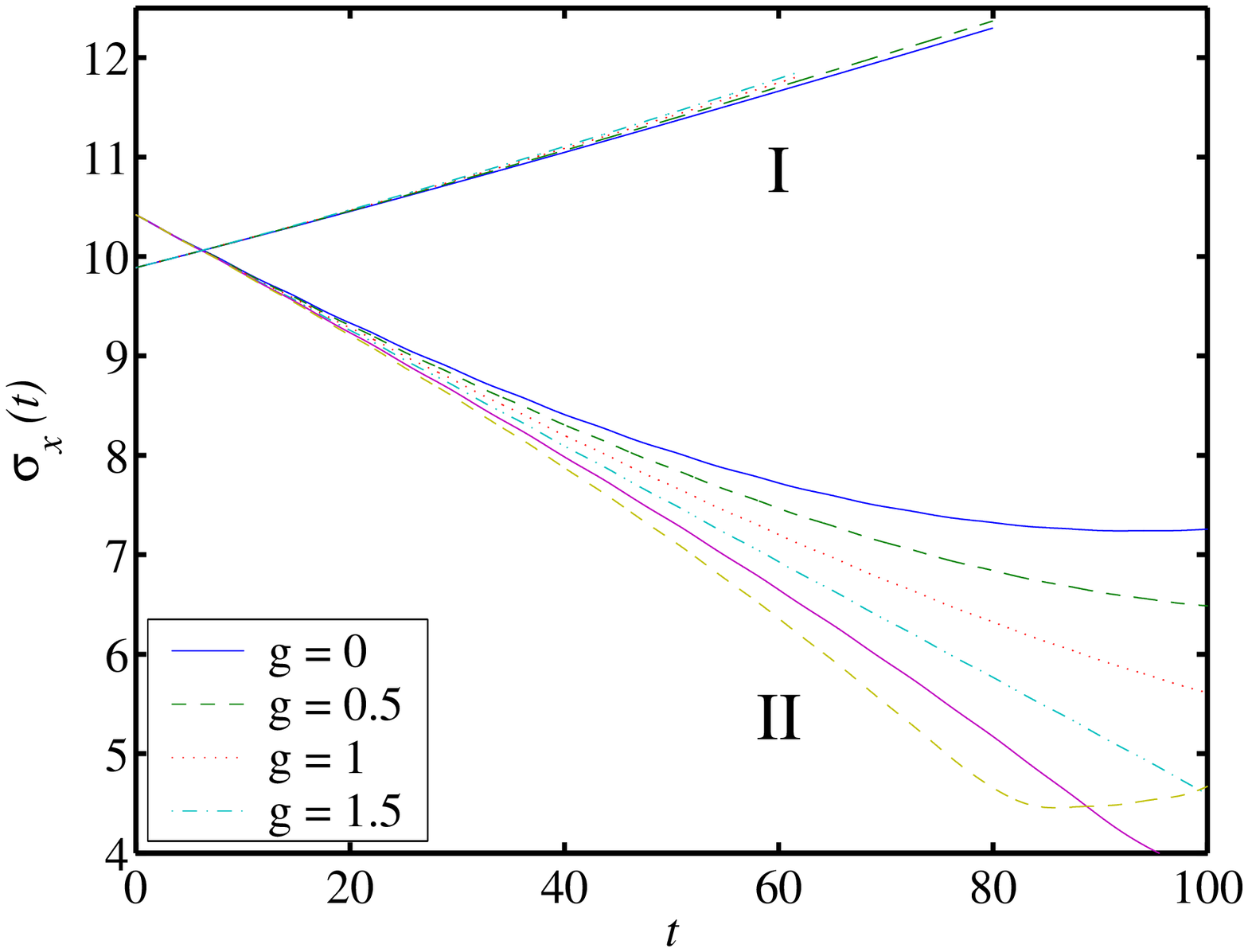}
		}
		\subfigure[]{		
		\label{fig:chi1_h}
		\includegraphics[width=2.75in]{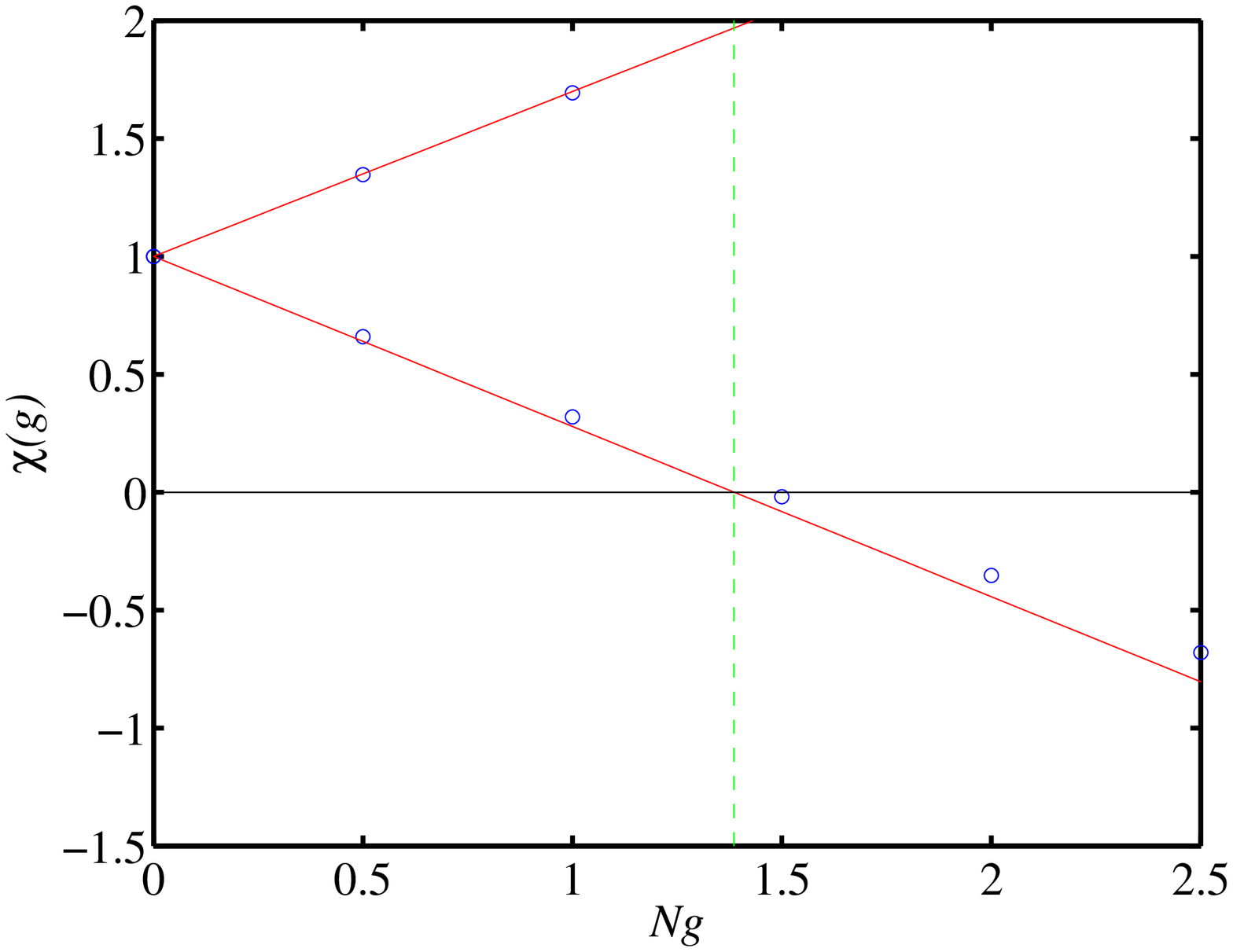}
		}
		\caption{(a) Dispersion for asymmetric honeycomb potential~(\ref{eq:pot-honey}) along high symmetry path. The green dashed line shows boundary between strongly bound and quasi-unbound states~\cite{ostrovskaya03}.
		(b) - $xx$, and $yy$ components of effective mass tensor along $k_y$-axis. 
		(c) - Dependence of wave packets dispersion on time after the lattice potential is introduced. The two sets of curves are for different points in Brillouin zone: (I) - point $\Gamma$, (II) - point K.
		(d) - Dispersive characteristic $\chi$ from~(\ref{eq:chi2}). The blue dots are obtained from fitting quadratic dispersion~(\ref{quadr_disp}) to continuous simulation data from panel (c), red line is expected behavior of $\chi$.}
		\label{fig:dispersion-meff-sigma-chi-h}
	\end{center}
\end{figure*}

\begin{figure}[floatfix]
	\begin{center}
		\subfigure[]{
		\label{fig:wfa}
		\includegraphics[width=2.75in]{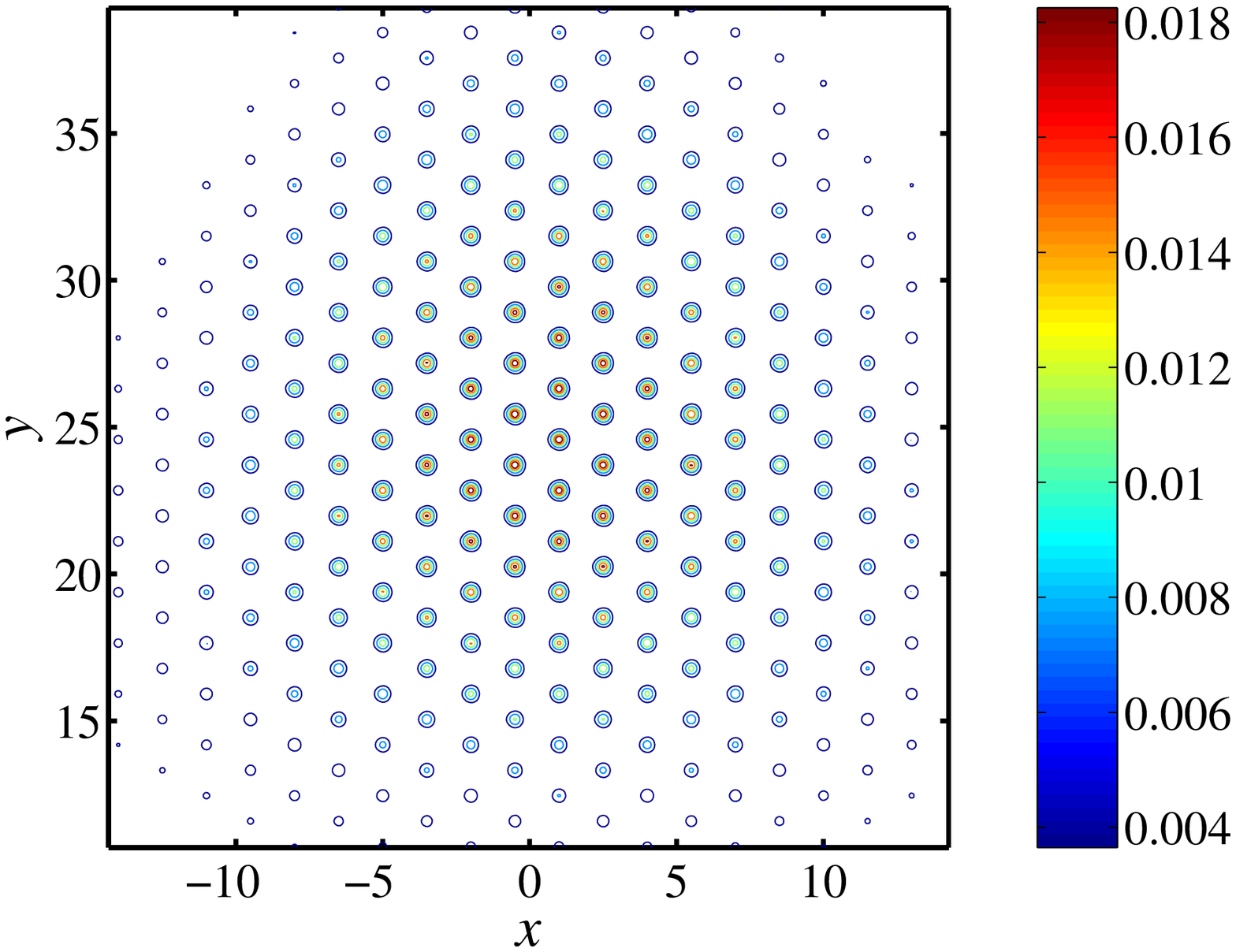}
		}
		\subfigure[]{		
		\label{fig:wfb}
		\includegraphics[width=2.75in]{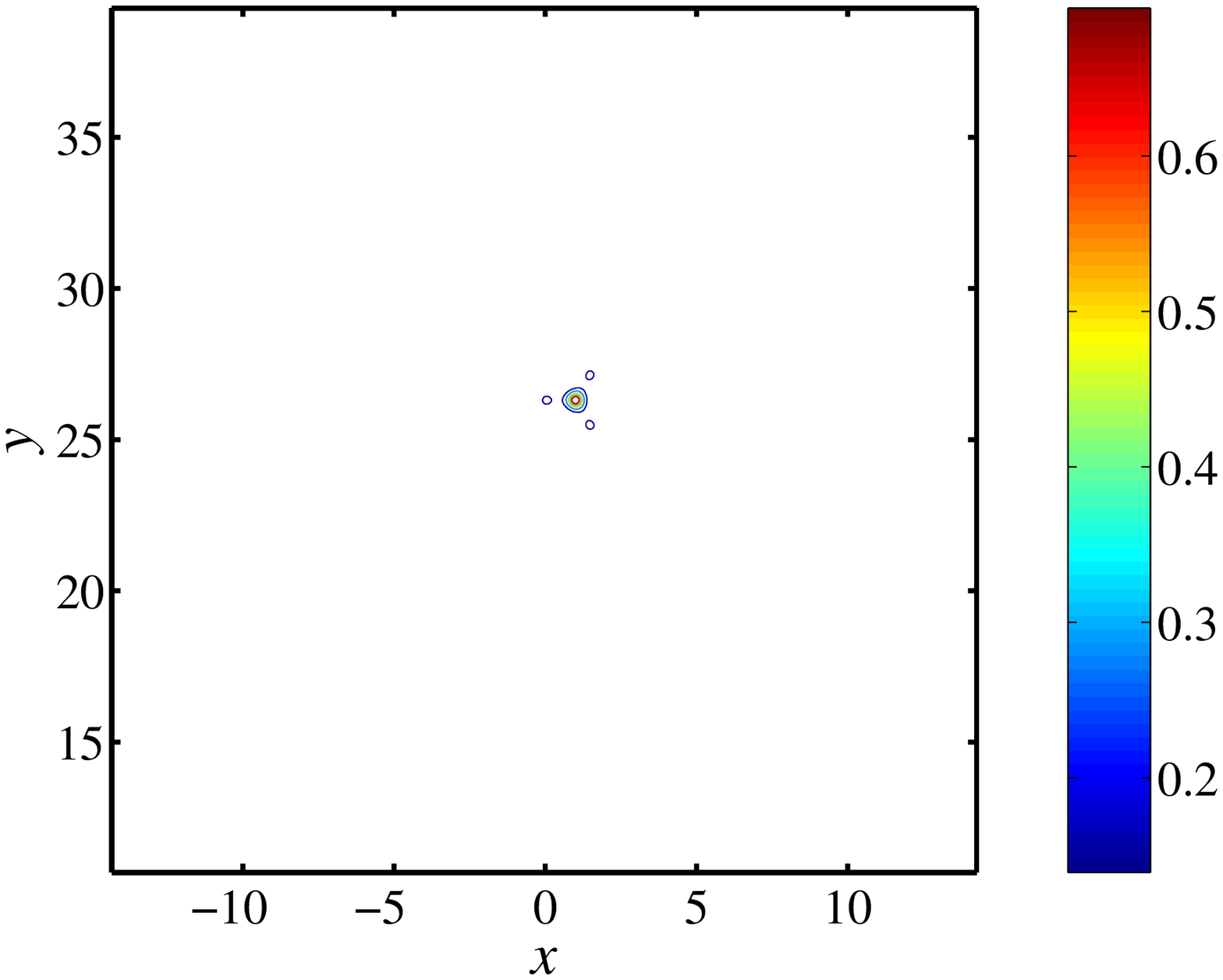}
		}
		\subfigure[]{
		\label{fig:wfc}
		\includegraphics[width=2.75in]{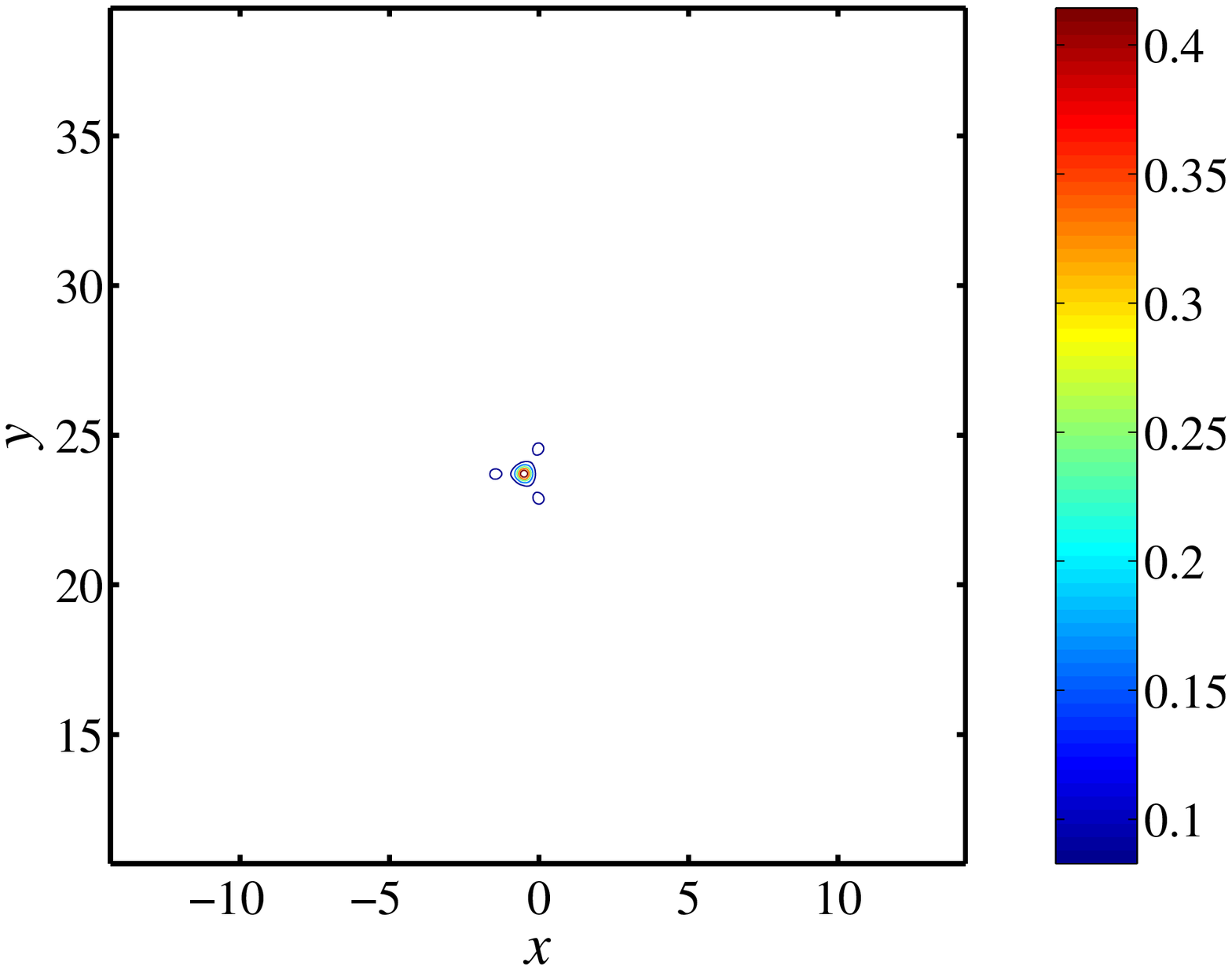}
		}
		\subfigure[]{		
		\label{fig:wfd}
		\includegraphics[width=2.75in]{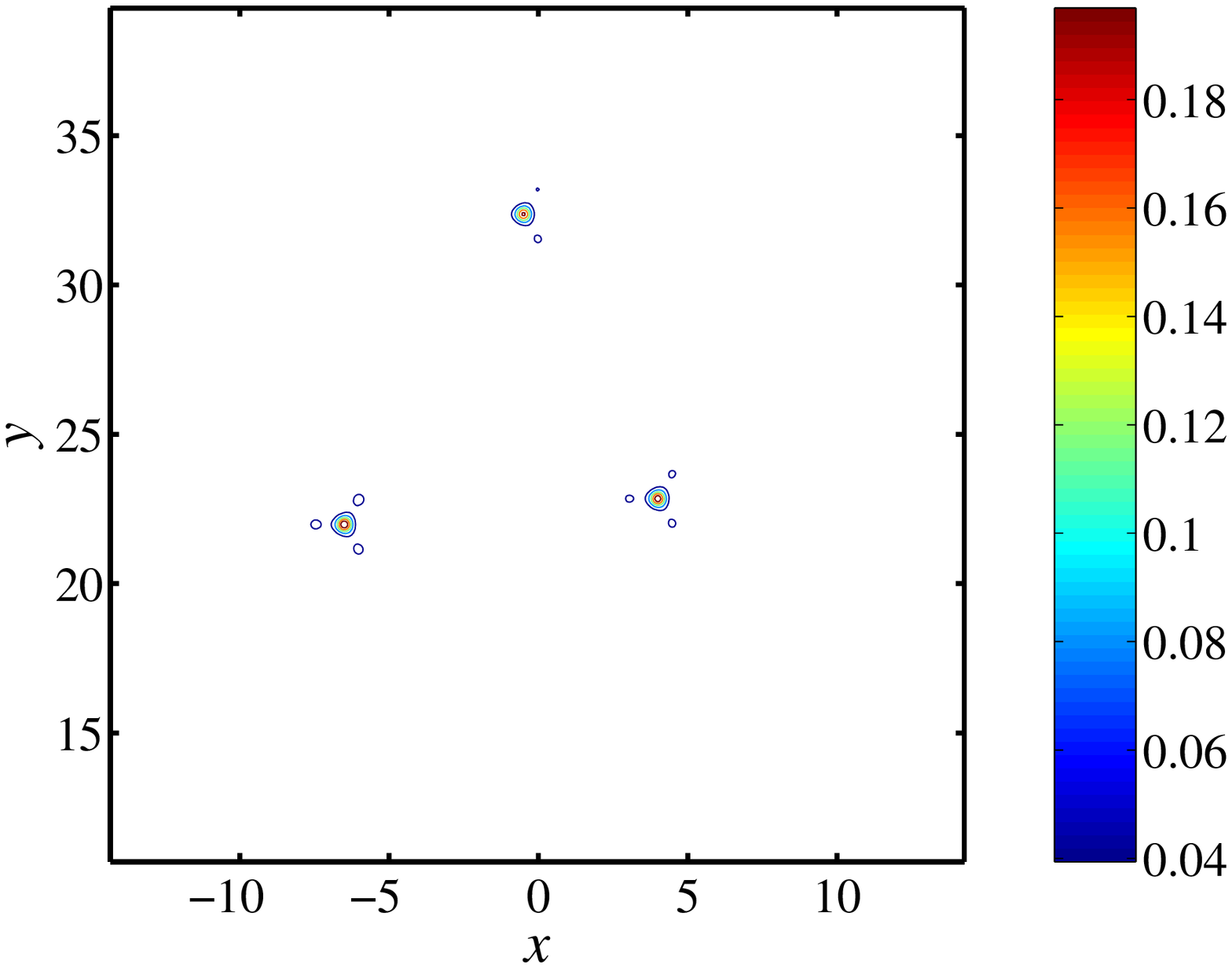}
		}
		\caption{Probability density of the wave packet that has been driven to K point of square lattice after expansion for $\Delta t_{\rm exp} = 200$ with different interactions: (a) - $N g_2 = 0.5$, (b) - $N g_2 = 2.5 $ (c) - $N g_2 = 5$ (d) - $N g_2 = 10$. As the interaction increases fraction of the wave function transferred to solitons decreases.}
		\label{fig:psis_h}
	\end{center}
\end{figure}

In our previous work~\cite{diener03} dedicated to study of quantum wave packet propagation in spatially asymmetric optical lattices above critical interaction strength, we have observed self collapse of the wave packet into localized modes. The effective mass concept provides an explanation of the phenomena and relates the critical interaction strength to other parameters of the problem.
Here, we describe briefly the model potential.
The potential used in~\cite{diener03} is
\begin{equation}
	\label{eq:pot-honey}
	V_{\rm lat} ({\bf r}) = \sum\limits_{i = 1}^6 {A_i \cos ({\bf k}_i  \cdot {\bf r} + \phi _i )},
\end{equation}
with the following parameters
\begin{equation}
\begin{array}{c}
 A_1  = A_3  = A_5  =  - 4, \\ 
 A_2  = A_4  = A_6  =  - 4.664, \\ 
 \phi _1  = \phi _3  = \phi _5  = 0, \\ 
 \phi _2  =  - \phi _4  = \phi _6  = 1.1. \\ 
 \end{array}
\end{equation}
It may be obtained either with interference of three pairs of counter propagating beams or by holographic techniques.
The band structure is shown in the Fig.~\ref{fig:dispersion_h}. Asymmetry splits the first band in the middle. In this case, the boundary between strongly bound and quasi-unbound states is located in the second gap. The effective mass is negative in both directions at K point (see Fig.~\ref{fig:meff_h}). We start with a Gaussian wave packet in free space which has spatial dispersion $\sigma _{x,y} = 10/\sqrt 2 \approx 7.071$. Since the size of the unit cell in this case is $\Omega = 3 \sqrt 3 /2$, the wave packet occupies several unit cells. The lattice potential is introduced in time $t _V = 120$ and for we accelerate it along $y$-axis with $\left| {\bf F} \right| = 0.05$. In this case conditions in ~(\ref{eq:tv})-(\ref{eq:f}) are fulfilled. Analysis of the expansion for different values of the interaction at the bottom of the first band ($\Gamma$ point) and the top of the first band (K point) is shown in the Fig.{\ref{fig:sigma1_h}. To compare results of the continuous simulation with the prediction of the~(\ref{eq:chi}) we calculated effective mass, $M_{\rm eff}$, from the band structure and the numerical factor $I$, that involves integrals of the Bloch wave functions, by integrating over one unit cell wave functions obtained by adiabatic evolution. Their values are $ M_{ {\rm eff}, \Gamma} = 1.7986$, $I_{\Gamma} = 0.9396$ at the point $\Gamma$ and $M_{\rm eff, K} \approx -0.8918$, $I_K\approx 1.9567$ at the point K. Probability density plots after expansion for $\Delta t_ {\rm exp} = 200$ at the K point are shown in the Fig.~\ref{fig:psis_h}. As interaction increased past critical interaction the lattice solitons are dynamically formed. The figure illustrates that the phenomena is also observed for a square lattice: the fraction of the wave function transferred to the localized modes decreases as the interaction strength is increased. The effective interaction experienced by each mode is such that corresponding chemical potential is close to the first band.

As we have seen in the numerical examples discussed above, if the nonlinearity is small enough, the wave packet collapses into a single DS. Clouds of ultracold atoms can be imaged non-destructively~\cite{andrews96}. Observation of a persistent atomic cloud localized to dimensions comparable to the wave length of the light forming the lattice and will be a clear signature of the DS formation. Observations of the predicted delocalizing transitions with a single DS prepared in a controlled fashion for a varying lattice depth is also an exciting possibility.

\section{Conclusion}
In this article we have shown how DS could be generated with a self-repulsive BEC in optical lattices by driving the wave packets to the points where the effective mass is negative, that leads to self collapse of the wave packet. As a result, part of the wave function is transferred to localized modes with chemical potential in between the first and second bands where DS with positive nonlinearity are supported. In 2D, this happens only when the critical nonlinearity is achieved. As the strength of interaction increases, the wave packet may excite several localized modes. We have observed that effective interaction of each mode is such that the corresponding chemical potential is located close to the top of the first band.

\section*{Acknowledgments}
We acknowledge support from the NSF, the R.A. Welch Foundation,
and discussions with J.Hanssen, K.C. Henderson, T.P. Meyrath, S.A. Moore and M.G. Raizen.

\end{document}